\documentclass[intlimits,twoside,a4paper]{article}

\usepackage{amsmath,amssymb}
\usepackage{graphicx}
\usepackage{wrapfig}

\usepackage[T2A]{fontenc}
\usepackage[cp1251]{inputenc}
\usepackage[eqsecnum]{cmpj2}

\issue{2012}{15}{4}{43702}

\doinumber{10.5488/CMP.15.43702}

\title[Distorted diamond spin chain at high fields and low temperatures]
{Semiquantitative theory for high-field low-temperature properties of a distorted \\ diamond spin chain}
\author[O.Derzhko, J.Richter, O.Krupnitska]{O. Derzhko\refaddr{label1,label2,label3},
                                            J. Richter\refaddr{label3},
                                            O. Krupnitska\refaddr{label2}}
\addresses{
\addr{label1} Institute for Condensed Matter Physics of the
              National Academy of Sciences of Ukraine,\\
              1 Svientsitskii St., 79011~Lviv, Ukraine
\addr{label2} Department for Theoretical Physics,
              Ivan Franko National University of Lviv,\\
              12 Drahomanov St., 79005~Lviv, Ukraine
\addr{label3} Institut f\"{u}r theoretische Physik,
              Universit\"{a}t Magdeburg,
              P.O. Box 4120, D--39016 Magdeburg, Germany
}

\authorcopyright{O.Derzhko, J.Richter, O.Krupnitska, 2012}
\date{Received June 10, 2012, in final form September 6, 2012}

\begin{document}

\maketitle

\begin{abstract}
We consider the antiferromagnetic Heisenberg model on a distorted diamond chain
and use the localized-magnon picture adapted to a distorted geometry
to discuss some of its high-field low-temperature properties.
More specifically, in our study we assume that the partition function for a slightly distorted geometry has the same form as for ideal geometry, though with slightly dispersive one-magnon energies.
We also discuss the relevance of such a description to azurite.
\keywords diamond spin chain, localized magnons, azurite
\pacs 75.10.Jm
\end{abstract}

\section{Introduction}
\label{sec1}
\setcounter{equation}{0}

The concept of localized magnons was introduced some time ago~\cite{lm(a),lm(b)}
and since then it has been successfully used
to examine the ground-state and low-temperature properties
of a wide class of spin models~\cite{localized_magnons(a),localized_magnons(b),localized_magnons(c),localized_magnons(d)}
(for a review see reference~\cite{fnt}).
Most of the calculations refer to the so-called ideal lattice geometry
which implies a completely dispersionless lowest-energy one-magnon band.
However, one cannot expect that such conditions occur in real-life materials
and, therefore, one has to go beyond the case of ideal geometry
dealing with a slightly dispersive lowest-energy one-magnon band, i.e., with almost localized magnons.
Although a systematic quantitative theory of almost localized magnons has not been elaborated so far,
it is quite in order to mention here
reference~\cite{distorted_ladder} that considers a distorted frustrated two-leg spin ladder
and
references~\cite{effective_xy(a),effective_xy(b)} where an effective Hamiltonian is obtained for a distorted diamond spin chain
[in the context of the magnetic compounds Cu$_3$Cl$_6$(H$_2$O)$_2\cdot$2H$_8$C$_4$SO$_2$ and Cu$_3$(CO$_3$)$_2$(OH)$_2$].

\begin{figure}
\begin{center}
\includegraphics[clip=on,width=65mm,angle=0]{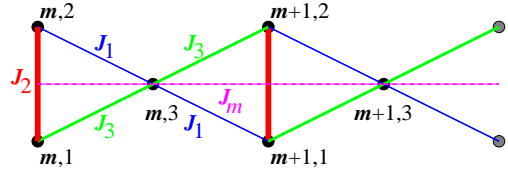}
\caption
{(Color online)
The distorted diamond spin chain considered in this paper.
The ideal diamond spin chain corresponds to $J_1=J_3$, $J_m=0$, $J_2>2J_1$.}
\label{fig01}
\end{center}
\end{figure}

A famous solid-state example of a model compound for a frustrated diamond Heisenberg spin-chain system
is the natural mineral azurite Cu$_3$(CO$_3$)$_2$(OH)$_2$~\cite{kikuchi(a),kikuchi(b),ah_azurite}.
For another possible experimental candidates see references~\cite{can1,can2}.
High-field magnetization curves for an azurite single crystal
have been measured below 4.2~K up to about 40~T, see references~\cite{kikuchi(a),kikuchi(b)}.
The magnetization curve has a clear plateau at one third of the saturation magnetization.
Moreover,
the magnetization curve has a very steep part (although not a perfect vertical jump)
between one third of the saturation magnetization and the saturation magnetization.
An appropriate magnetic model for azurite~\cite{kikuchi(a),kikuchi(b),ah_azurite}
is a spin-1/2 distorted Heisenberg diamond chain
with four different antiferromagnetic exchange constants $J_1$, $J_2$, $J_3$, and $J_m$,
see figure~\ref{fig01} and sections~\ref{sec2}, \ref{sec4}.
The ideal geometry supporting the localized-magnon states occurs for the case $J_1=J_3$, $J_m=0$, $J_2>2J_1$.
However,
the set of exchange constants
obtained from the first-principle density functional computations reads~\cite{ah_azurite}
\begin{eqnarray}
\label{1.01}
J_1=15.51\ {\text{K}}, \quad J_2=33\ {\text{K}}, \quad J_3=6.93\ {\text{K}}, \quad J_m=4.62\ {\text{K}}.
\end{eqnarray}
Although the parameter set (\ref{1.01}) does not satisfy the ideal geometry conditions,
it is not very far from an ideal set.
This statement is supported by the measured magnetization curve
that resembles the one predicted by a localized-magnon picture,
see~\cite{localized_magnons(a),localized_magnons(b),localized_magnons(c),localized_magnons(d),fnt}.
Therefore,
one may expect that an appropriate modification of the localized-magnon picture
would be capable of describing the high-field low-temperature properties of azurite.

It should be mentioned here that the diamond spin chain with $J_1=J_3$, $J_m=0$
is a representative of the models with local conservation laws,
cf., e.g., references~\cite{gelfand91,niggemann(a),niggemann(b),ivanov(a),ivanov(b)}.
Local conservation laws provide a very special mechanism for trapping the magnons.
Therefore, the ideal diamond chain cannot be considered as a generic spin model with localized magnons.
On the other hand,
this is probably the most suitable well known solid-state realization of a localized-magnon system.

Bearing in mind this motivation,
in the present paper we discuss
one simple route to the high-field low-temperature thermodynamics of a distorted diamond spin chain
which is based on the localized-magnon picture.
After recalling in section~\ref{sec2} the basic points of the standard consideration
which is valid for ideal geometry~\cite{localized_magnons(a),localized_magnons(b),localized_magnons(c),localized_magnons(d),fnt},
we introduce in section~\ref{sec3} a plausible form of the partition function
of the distorted diamond spin chain for small deviations from ideal geometry
and calculate thermodynamic quantities in this case.
In section~\ref{sec4} we apply this consideration to azurite.
We conclude in section~\ref{sec5} with a brief summary and prospects for further studies.

\section{Localized magnons on an ideal diamond chain}
\label{sec2}
\setcounter{equation}{0}

In our study we consider the spin-1/2 Heisenberg antiferromagnet with the Hamiltonian
\begin{eqnarray}
\label{2.01}
H=\sum_{(ij)}J_{ij}{\bf{s}}_i \cdot{\bf{s}}_j-hS^z,
\qquad
S^z=\sum_is_i^z
\end{eqnarray}
on a $N$-site diamond chain~\cite{niggemann(a),niggemann(b),diamond(a),diamond(b),diamond(c),diamond(d),diamond(e)}, see figure~\ref{fig01}.
We use standard notations in equation~(\ref{2.01}) and imply periodic boundary conditions.
It is convenient to label the lattice sites with a pair of indeces,
where the first number enumerates the cells ($m=1,\ldots,{\cal{N}}=N/3$)
and the second one enumerates the position of the site within the cell,
see figure~\ref{fig01}.
Therefore,
spin Hamiltonian~(\ref{2.01}) on the distorted diamond-chain lattice reads
\begin{eqnarray}
\label{2.02}
H&=&\sum_{m=1}^{\cal{N}}
\big[
J_2{\bf{s}}_{m,1}\cdot{\bf{s}}_{m,2}
+J_3{\bf{s}}_{m,1}\cdot{\bf{s}}_{m,3}+J_1{\bf{s}}_{m,2}\cdot{\bf{s}}_{m,3}
+J_1{\bf{s}}_{m,3}\cdot{\bf{s}}_{m+1,1}+J_3{\bf{s}}_{m,3}\cdot{\bf{s}}_{m+1,2}
\nonumber\\
&&
+J_m{\bf{s}}_{m,3}\cdot{\bf{s}}_{m+1,3}
-h\left(s_{m,1}^z+s_{m,2}^z+s_{m,3}^z\right)
\big].
\end{eqnarray}
Note that the spin Hamiltonian commutes with $S^z$.
Therefore, we may consider the subspaces with different values of $S^z=N/2,N/2-1,\ldots$ separately.
Moreover, we may assume for brevity at first $h=0$ and then trivially add the contribution of the Zeeman term.

We begin with the ideal geometry case.
If $J_1=J_3$, $J_m=0$,
the one-magnon (i.e., $S^z=3{\cal{N}}/2-1$) energy bands
$\varepsilon_i(\kappa)$,
$i=1,2,3$,
$\kappa=2\pi n/{\cal{N}}$,
$n=-{\cal{N}}/2,-{\cal{N}}/2+1,\ldots,{\cal{N}}/2-1$
(we assume without loss of generality that ${\cal{N}}$ is even)
follow from the equation
 \begin{eqnarray}
\label{2.03}
{\mathrm{det}}\left(
\begin{array}{ccc}
-J_1-\frac{J_2}{2}-\varepsilon_i(\kappa) & \frac{J_2}{2} & \frac{J_1}{2}\left(1+\re^{\ri\kappa}\right) \\
\frac{J_2}{2} & -J_1-\frac{J_2}{2}-\varepsilon_i(\kappa) & \frac{J_1}{2}\left(1+\re^{\ri\kappa}\right) \\
\frac{J_1}{2}\left(1+\re^{-\ri\kappa}\right) & \frac{J_1}{2}\left(1+\re^{-\ri\kappa}\right) & -2J_1-\varepsilon_i(\kappa)
\end{array}
\right)=0
\end{eqnarray}
and have the form
\begin{eqnarray}
\label{2.04}
\varepsilon_1(\kappa)=-J_1-J_2,
\qquad
\varepsilon_{2,3}(\kappa)=-\frac{3J_1}{2}\mp\frac{J_1}{2}\sqrt{5+4\cos\kappa} \,.
\end{eqnarray}
For $J_1/J_2<1/2$
(from now on we assume that this inequality holds)
the flat band $\varepsilon_1(\kappa)=\varepsilon_1$ becomes the lowest-energy one.
The states from the flat band can be visualized as singlets located on the vertical bonds $J_2$,
see figure~\ref{fig01}.

Many-magnon (i.e., $S^z=3{\cal{N}}/2-2,\ldots,3{\cal{N}}/2-{\cal{N}}$) ground states can be constructed
by filling the vertical bonds (traps) by magnons.
These independent localized-magnon states dominate the low-temperature properties of the spin system
in a magnetic field $h$ around the saturation value $h_{\rm{sat}}=-\varepsilon_1=J_1+J_2$.
The partition function of the diamond spin chain in this regime reads~\cite{localized_magnons(a),localized_magnons(b),localized_magnons(c),localized_magnons(d),fnt}
\begin{eqnarray}
\label{2.05}
Z(T,h,N)=\exp\left({-\frac{E_{{\rm{FM}}}-\frac{h}{2}N}{T}}\right)
\sum_{n=0}^{\cal{N}}g_{\cal{N}}(n)\re^{n\left(h_{\rm{sat}}-h\right)/{T}}
\end{eqnarray}
(we set $k_{\rm{B}}=1$)
with
$E_{{\rm{FM}}}=(J_2/4+J_1){\cal{N}}$
and
$g_{\cal{N}}(n)={\cal{N}}!/(n!({\cal{N}}-n)!)$.
Therefore,
\begin{eqnarray}
\label{2.06}
Z(T,h,N)=\exp\left(-\frac{E_{{\rm{FM}}}-\frac{h}{2}N}{T}\right)
\left[1+\re^{\left(h_{\rm{sat}}-h\right)/{T}}\right]^{\cal{N}}
\end{eqnarray}
and the free energy per cell reads
\begin{eqnarray}
\label{2.07}
f(T,h)=\lim_{{\cal{N}}\to\infty}\frac{-T\ln Z(T,h,N)}{{\cal{N}}}
=\frac{E_{{\rm{FM}}}-\frac{h}{2}N}{{\cal{N}}}
-T\ln\left[1+\re^{\left(h_{\rm{sat}}-h\right)/{T}}\right].
\end{eqnarray}
The magnetization per cell can be obtained from equation~(\ref{2.07}) according to the standard relation
\begin{eqnarray}
\label{2.08}
m(T,h)=-\frac{\partial f(T,h)}{\partial h}
=\frac{3}{2}
- \frac{\re^{\left(h_{\rm{sat}}-h\right)/{T}}}{1+\re^{\left(h_{\rm{sat}}-h\right)/{T}}}\, .
\end{eqnarray}
Some further calculations of the high-field low-temperature thermodynamic quantities can be found in~\cite{localized_magnons(a),localized_magnons(b),localized_magnons(c),localized_magnons(d),fnt}.

\section{Almost localized magnons on a distorted diamond chain}
\label{sec3}
\setcounter{equation}{0}

Now we assume a small ``distortion'' $\vert J_1-J_3\vert/J_2 \ll 1$, $J_m/J_2\ll 1$.
Our aim is to construct an effective description
of the low-temperature thermodynamics of the distorted diamond spin chain~(\ref{2.02}) around $h_1=(J_1+J_3)/2+J_2$.

We begin with the one-magnon energy bands $\varepsilon_i(\kappa)$,
which follows now from the equation
\begin{eqnarray}
\label{3.01}
\mbox{det}\!\!\left(\!\!
\begin{array}{ccc}
-J-\frac{J_2}{2}-\varepsilon_i(\kappa) & \frac{J_2}{2} & \frac{J}{2}\left(1+\re^{\ri\kappa}\right)-\frac{\delta}{2}\left(1-\re^{\ri\kappa}\right) \\
\frac{J_2}{2} & -J-\frac{J_2}{2}-\varepsilon_i(\kappa) & \frac{J}{2}\left(1+\re^{\ri\kappa}\right)+\frac{\delta}{2}\left(1-\re^{\ri\kappa}\right) \\
\frac{J}{2}\left(1+\re^{-\ri\kappa}\right)-\frac{\delta}{2}\left(1-\re^{-\ri\kappa}\right) & \frac{J}{2}\left(1+\re^{-\ri\kappa}\right)+\frac{\delta}{2}\left(1-\re^{-\ri\kappa}\right) & -2J-J_m(1-\cos\kappa)-\varepsilon_i(\kappa)
\end{array}\!\!
\right)\!\!=0. \nonumber
\end{eqnarray}\vspace{-7mm}
\begin{eqnarray}
\end{eqnarray}
Here, we have introduced the following notations
\begin{eqnarray}
\label{3.02}
J=\frac{J_1+J_3}{2}\, ,
\qquad
\delta=\frac{J_1-J_3}{2}\, .
\end{eqnarray}
Equation~(\ref{3.01}) yields a cubic equation for $\varepsilon_i(\kappa)$,
\begin{eqnarray}
\label{3.03}
&&-\left[J+\frac{J_2}{2}+\varepsilon_i(\kappa)\right]^2
\left[2J+2J_m\sin^2\frac{\kappa}{2}+\varepsilon_i(\kappa)\right]
+J_2\left(J^2\cos^2\frac{\kappa}{2}-\delta^2\sin^2\frac{\kappa}{2}\right)
\\
&&+\left[2J+J_2+2\varepsilon_i(\kappa)\right]\left(J^2\cos^2\frac{\kappa}{2}+\delta^2\sin^2\frac{\kappa}{2}\right)
+\frac{J_2^2}{4}\left[2J+2J_m\sin^2\frac{\kappa}{2}+\varepsilon_i(\kappa)\right]=0.\nonumber
\end{eqnarray}

In the ideal geometry limit $J=J_1$, $\delta=0$, $J_m=0$,
equation~(\ref{3.03}) becomes
\begin{eqnarray}
\label{3.04}
-\left[J_1+\frac{J_2}{2}+\varepsilon_i(\kappa)\right]^2\left[2J_1+\varepsilon_i(\kappa)\right]
+2\left[J_1+J_2+\varepsilon_i(\kappa)\right]J_1^2\cos^2\frac{\kappa}{2}
+\frac{J_2^2}{4}\left[2J_1+\varepsilon_i(\kappa)\right]=0.
\end{eqnarray}
Obviously, $\varepsilon_1(\kappa)=\varepsilon_1=-J_1-J_2$ (\ref{2.04}) satisfies equation~(\ref{3.04}).

We pass to the distorted case assuming $\delta/J_2$ to be small.
Inserting
\begin{eqnarray}
\label{3.05}
\varepsilon_1(\kappa)=-J-J_2+\varepsilon_1^{(2)}\delta^2+\ldots
\end{eqnarray}
into equation~(\ref{3.03})
and collecting the terms of the order $\delta^2$ we find
\begin{eqnarray}
\label{3.06}
\varepsilon_1^{(2)}
=\frac{2J_2\sin^2\frac{\kappa}{2}}{J_2(J-J_2)+2J^2\cos^2\frac{\kappa}{2}+2J_2J_m\sin^2\frac{\kappa}{2}}\, .
\end{eqnarray}
Thus, up to the terms of the order $\delta^2$, we have
\begin{eqnarray}
\label{3.07}
\varepsilon_1(\kappa)
\!\!\!\!&=&\!\! \!\!
-J-J_2
+\frac{1}{1-\frac{J}{J_2}-(1+\cos\kappa)\frac{J^2}{J_2^2} - (1-\cos\kappa)\frac{J_m}{J_2}}
\frac{(J_1-J_3)^2}{4J_2}(-1+\cos\kappa)
+\ldots
\\
&=&\!\!\!\!
-h_1
+\frac{1}{1-\frac{J}{J_2}-\frac{J_m}{J_2}-\frac{J^2}{J_2^2}+\left(\frac{J_m}{J_2}-\frac{J^2}{J_2^2}\right)\cos\kappa}
\frac{(J_1-J_3)^2}{4J_2}(-1+\cos\kappa)
+\ldots \, .\nonumber
\end{eqnarray}
Clearly, the flat band $\varepsilon_1$ becomes dispersive
if $\delta \ne 0$.
Interestingly,
as it follows from equation~(\ref{3.07}) and can be expected from the coupling geometry in figure~\ref{fig01},
the exchange coupling $J_m$ affects the flat band only if $J_1\ne J_3$
(i.e., $J_m$ cannot spoil the flat band alone).
Furthermore,
assuming in addition that $J/J_2$, $J_m/J_2$ are also small,
we can rewrite equation~(\ref{3.07}) in the form
\begin{eqnarray}
\label{3.08}
\varepsilon_1(\kappa)
\approx
-h_1-\frac{(J_1-J_3)^2}{4J_2}
+\frac{(J_1-J_3)^2}{4J_2}\cos\kappa \, .
\end{eqnarray}

\begin{figure}[!b]
\begin{center}
\includegraphics[clip=on,width=0.49\textwidth,angle=0]{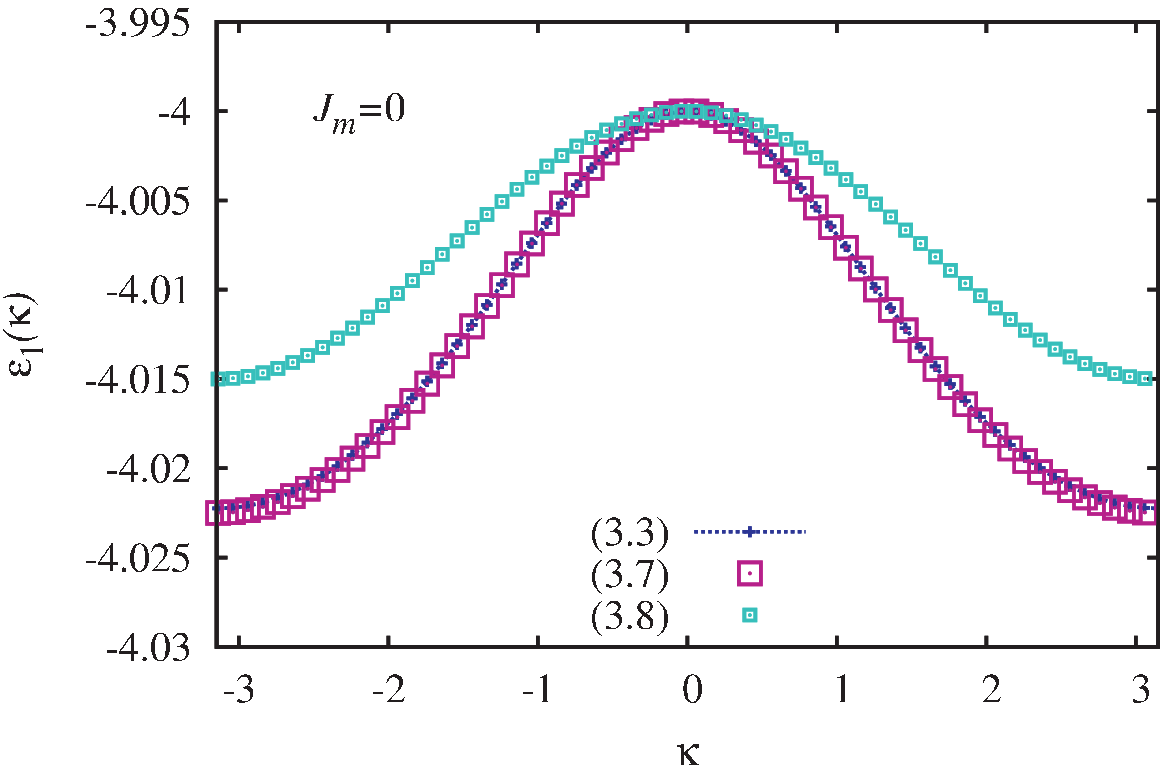}\hfill   
\includegraphics[clip=on,width=0.49\textwidth,angle=0]{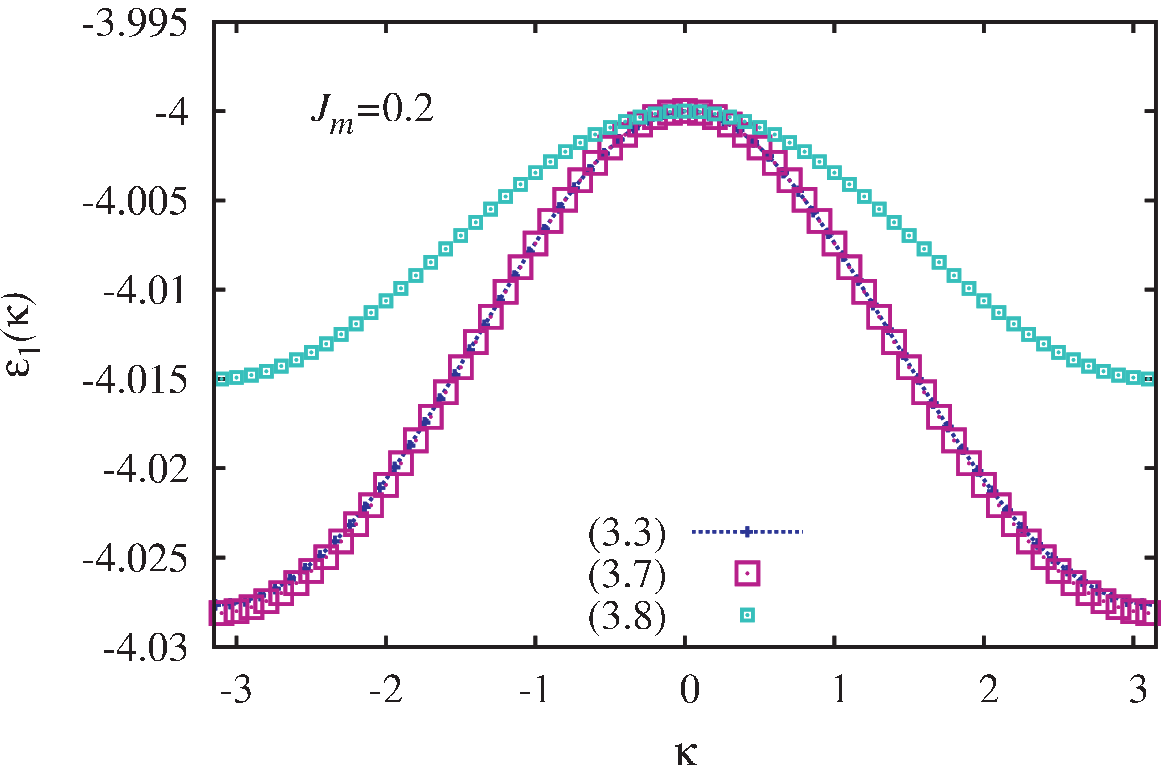}
\caption
{(Color online)
$\varepsilon_1(\kappa)$ obtained from
equation~(\ref{3.03}) (line),
equation~(\ref{3.07}) (large squares),
and
equation~(\ref{3.08}) (small squares)
for the set of parameters $J_1=0.85$, $J_2=3$, $J_3=1.15$,
$J_m=0$ (left hand panel),
and
$J_m=0.2$ (right hand panel).}
\label{fig02}
\end{center}
\end{figure}

In figure~\ref{fig02} we compare $\varepsilon_1(\kappa)$
obtained from equations~(\ref{3.03}), (\ref{3.07}), and (\ref{3.08})
for two particular sets of parameters $J_1=0.85$, $J_2=3$, $J_3=1.15$, $J_m=0$ and
$J_m=0.2$, cf. equation~(\ref{1.01}).

Now we return to equation~(\ref{2.06})
which takes into account the contribution of the many-magnon states  to the thermodynamics
for ideal geometry when the lowest-energy one-magnon band is completely flat.
Evidently, equation~(\ref{2.06}) can be considered as
\begin{eqnarray}
\label{3.09}
Z(T,h,N)=\exp\left(-\frac{E_{{\rm{FM}}}-\frac{h}{2}N}{T}\right)
\prod_\kappa\left\{1+\re^{\left[-\varepsilon_1(\kappa)-h\right]/{T}}\right\},
\end{eqnarray}
where $-\varepsilon_1(\kappa)=-\varepsilon_1=h_{\rm{sat}}$ for the flat-band case.
Our basic assumption for the distorted case is as follows:
We assume that the partition function for a slightly distorted diamond spin chain still has the form given in equation~(\ref{3.09}), though with the one-magnon energies given in equation~(\ref{3.07}).
Preserving the structure of the partition function, we adopt the hard-monomer rule,
though facing now the hard monomers with slightly dispersive energies.
Thus, we have
\begin{eqnarray}
\label{3.10}
Z(T,h,N)=\exp\left(-\frac{E_{{\rm{FM}}}-\frac{h}{2}N}{T}\right)
\prod_\kappa\left\{1+\re^{\left[{-\varepsilon_1(\kappa)-h}\right]/{T}}\right\}
\end{eqnarray}
with $\varepsilon_1(\kappa)$ given in equation~(\ref{3.07}).
As a result, the free energy per cell reads
\begin{eqnarray}
\label{3.11}
f(T,h)=\lim_{{\cal{N}}\to\infty}\frac{-T\ln Z(T,h,N)}{{\cal{N}}}
=\frac{E_{{\rm{FM}}}-\frac{h}{2}N}{{\cal{N}}}
-\frac{T}{2\pi}\int_{ -\pi}^{\pi}\rd\kappa\ln\left\{1+\re^{\left[{-\varepsilon_1(\kappa)-h}\right]/{T}}\right\}
\end{eqnarray}
with $\varepsilon_1(\kappa)$ given in equation~(\ref{3.07}).
A further improved (but more complicated) result will be obtained
if one utilizes for $\varepsilon_1(\kappa)$ the corresponding solution of cubic equation (\ref{3.03}),
see section~\ref{sec4}.

It is worth noting
that the second term in equation~(\ref{3.11}) with $\varepsilon_1(\kappa)$ given by equation~(\ref{3.08})
corresponds to the free energy per site
(up to an unimportant constant ${\sf{h}}/2$)
of the spin-1/2 $XX$ chain in a transverse field~\cite{lieb(a),lieb(b),lieb(c)}
defined by the Hamiltonian
\begin{eqnarray}
\label{3.12}
{\sf{H}}=-{\sf{h}}\sum_{m=1}^{\cal{N}}\tau_m^z+{\sf{J}}\sum_{m=1}^{\cal{N}}
\left(\tau_m^x\tau_{m+1}^x+\tau_m^y\tau_{m+1}^y\right)
\end{eqnarray}
with
\begin{eqnarray}
\label{3.13}
{\sf{h}}= -h+h_1+\frac{(J_1-J_3)^2}{4J_2}\,,
\qquad
{\sf{J}}=\frac{(J_1-J_3)^2}{4J_2}\,.
\end{eqnarray}
This finding can be compared with the effective Hamiltonian for a distorted diamond spin chain
derived in references~\cite{effective_xy(a),effective_xy(b)}
within the second-order perturbation theory in $\vert J_1\vert/J_2$, $\vert J_3\vert/J_2$.
This  effective Hamiltonian also corresponds to the spin-1/2 $XX$ chain in a transverse field (\ref{3.12})
with the parameters
\begin{eqnarray}
\label{3.14}
{\sf{h}}=h-h_1-\frac{(J_1-J_3)^2}{4J_2}\,,
\qquad
{\sf{J}}=\frac{(J_1-J_3)^2}{4J_2}\,,
\end{eqnarray}
see equations~(7) and (8) of reference~\cite{effective_xy(b)}.

Furthermore,
within the adopted approach it is easy to obtain the magnetization.
Using~(\ref{3.11}) one gets
\begin{eqnarray}
\label{3.15}
m(T,h)
=
\frac{3}{2}-\frac{1}{2\pi}\int_{-\pi}^{\pi}\rd\kappa
\frac{\re^{\left[{-\varepsilon_1(\kappa)-h}\right]/{T}}}{1+\re^{\left[-\varepsilon_1(\kappa)-h\right]/{T}}}
=
1-\frac{1}{4\pi}\int_{-\pi}^{\pi}\rd\kappa\tanh\frac{-\varepsilon_1(\kappa)-h}{2T}\,.
\end{eqnarray}
Here, $\varepsilon_1(\kappa)$ is given by equation~(\ref{3.07})
[or by the corresponding solution of cubic equation (\ref{3.03})].
If one assumes for $\varepsilon_1(\kappa)$ the simpler formula (\ref{3.08}),
the second term in the r.h.s. of equation~(\ref{3.15}) corresponds to the behavior
of the (transverse) magnetization of the spin-1/2 $XX$ chain in a transverse field.

\begin{wrapfigure}{i}{0.485\textwidth}
\vspace{-5mm}
\begin{center}
\includegraphics[clip=on,width=0.48\textwidth,angle=0]{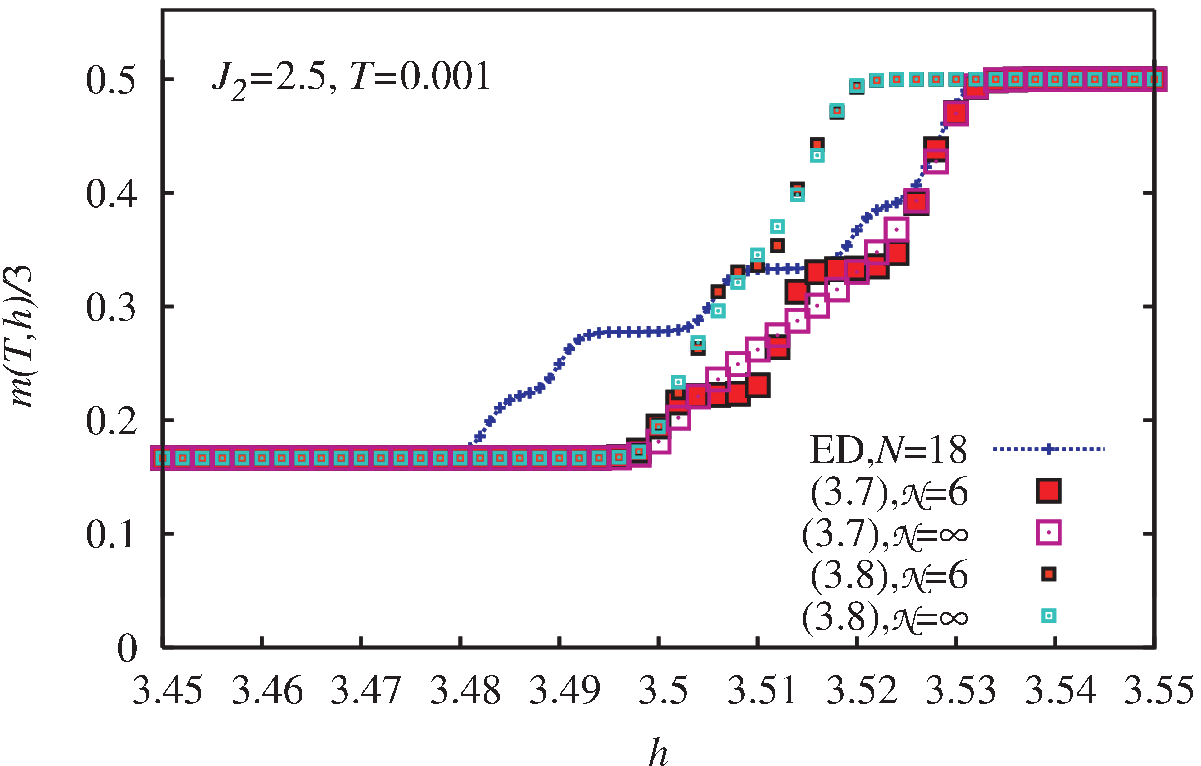}
\vspace{1mm}
\\
\includegraphics[clip=on,width=0.48\textwidth,angle=0]{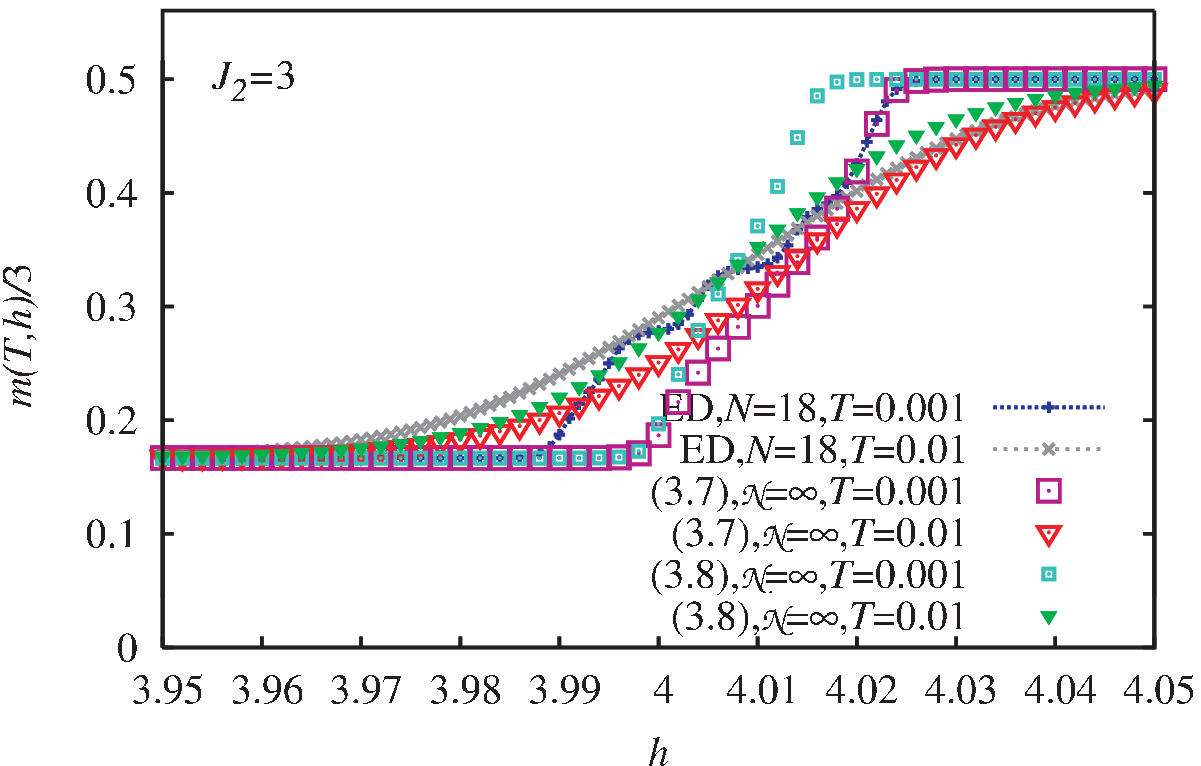}
\vspace{1mm}
\\
\includegraphics[clip=on,width=0.48\textwidth,angle=0]{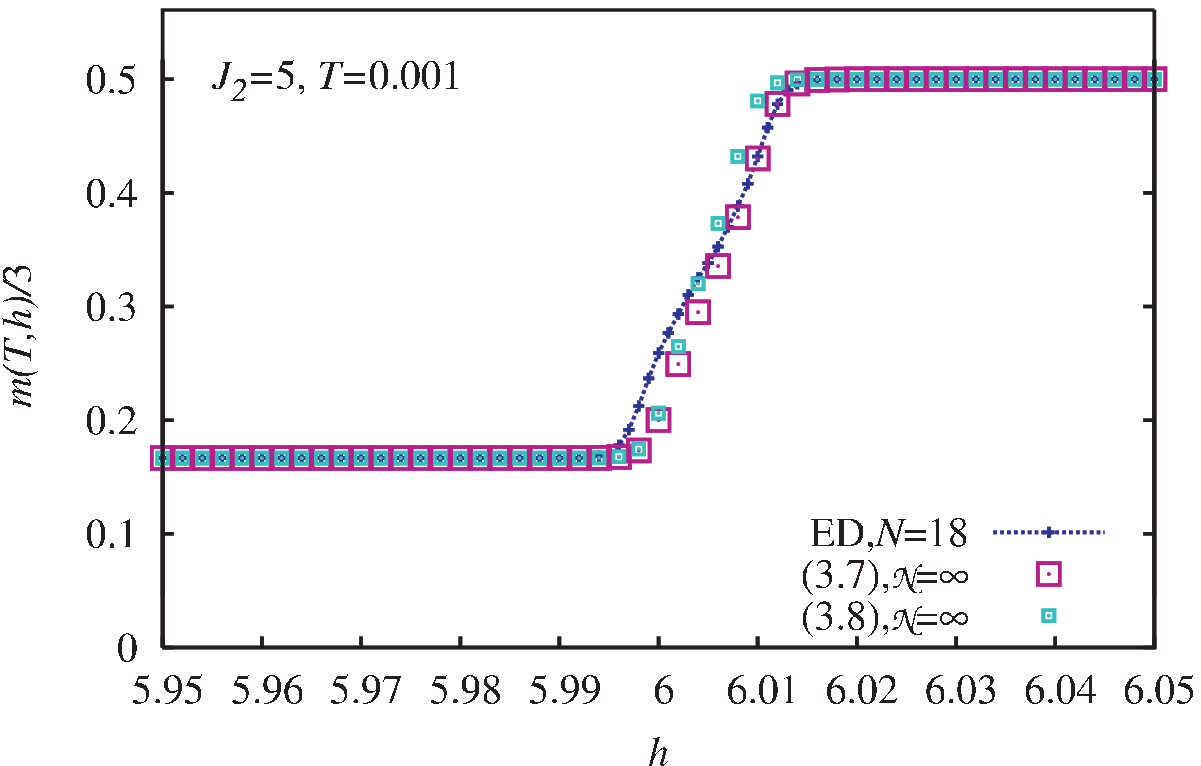}
\caption
{(Color online)
Magnetization curve $m(T,h)/3$ vs $h$ for the distorted diamond chain
[$J_1=0.85$,
$J_2=2.5$ (upper panel), $J_2=3$ (middle panel), $J_2=5$ (lower panel),
$J_3=1.15$,
$J_m=0$]
at $T=0.001$ (thick lines, squares) and $T=0.01$ (thin line, triangles).
Exact diagonalization data for $N=18$: lines;
approximate analytical theory which uses equation~(\ref{3.07}):
large empty (${\cal{N}}\to\infty$) and filled (${\cal{N}}=6$) symbols;
approximate analytical theory which uses equation~(\ref{3.08}):
small empty (${\cal{N}}\to\infty$) and filled (${\cal{N}}=6$) symbols.}
\label{fig03}
\end{center}
\vspace{-1.5cm}
\end{wrapfigure}

In figure~\ref{fig03}
we compare the exact diagonalization data with our predictions from the approximate analytical theory.
Exact diagonalization data refer to finite systems of $N=18$ sites.
The ground-state magnetization curve for finite systems consists of the steps
which become smeared out as the temperature increases.
Although analytical predictions according to equation~(\ref{3.15}) refer to thermodynamically large systems with ${\cal{N}}\to\infty$,
we may reproduce the finite-${\cal{N}}$ magnetization
replacing the integral by the sum,
i.e., $\int_{-\pi}^{\pi}\rd\kappa(\ldots)/(2\pi) \to \sum_\kappa(\ldots)/{\cal{N}}$.
Comparing the exact diagonalization data and approximate analytical calculations, one concludes the following.
Equation~(\ref{3.08})
(it corresponds to the effective spin-1/2 $XX$ chain in a transverse field introduced in references~\cite{effective_xy(a),effective_xy(b)})
only qualitatively reproduces the exact diagonalization data for $\vert J_1-J_3\vert/J_2=0.12,\,0.10$;
the value of the saturation field is underestimated, the end field for the 1/3 plateau equals $h_1$ and is overestimated.
Equation~(\ref{3.07}) works much better for $\vert J_1-J_3\vert/J_2=0.12,\,0.10$
and provides a good agreement with the exact diagonalization data above and just below the saturation field.
However, the end field for the 1/3 plateau again equals $h_1$
and around $h_1$ both approximations~(\ref{3.07}) and~(\ref{3.08}) exhibit similar shortcomings.
For a smaller value of $\vert J_1-J_3\vert/J_2=0.06$,
the agreement between both approximations and with the exact diagonalization data becomes better.
This is not surprising since equation~(\ref{3.08}) corresponds to the second-order perturbation theory in $\vert J_1\vert/J_2$, $\vert J_3\vert/J_2$,
see references~\cite{effective_xy(a),effective_xy(b)}.

We conclude this section by making some general remarks concerning the suggested approximation (\ref{3.10}) based on the discussed results.
Apparently, this educated ansatz which originates from the localized-magnon theory works well when the number of magnons is small
(i.e., around the saturation field)
provided the one-magnon energies are reproduced correctly
[compare $\varepsilon_1(\kappa)$
obtained from equations~(\ref{3.03}), (\ref{3.07}), and~(\ref{3.08})
and shown in the left hand panel of figure~\ref{fig02}].
If the number of magnons becomes large
(i.e., when approaching the end field for the 1/3 plateau)
a simple hard-core rule fails to describe the system
since the incompletely localized magnons may exhibit more complicated interactions.

\section{Magnetization curves for azurite}
\label{sec4}
\setcounter{equation}{0}

The natural mineral azurite Cu$_3$(CO$_3$)$_2$(OH)$_2$ has been a subject of intensive experimental and theoretical studies recently.
After the discovery of a plateau at 1/3 of the saturation value at the low-temperature magnetization curve~\cite{kikuchi(a),kikuchi(b)},
there were other experiments concerning the magnetic properties of azurite,
e.g., measurements of the magnetic susceptibility,
the specific heat,
the structure of the 1/3 plateau determined by nuclear magnetic resonance,
inelastic neutron scattering on the 1/3 plateau etc.,
see reference~\cite{effective_xy(b)} and references therein.
Applying different theoretical tools,
it was demonstrated that a generalized diamond spin chain is consistent with these experiments
and thus, azurite Cu$_3$(CO$_3$)$_2$(OH)$_2$ may be viewed as a model substance for a frustrated diamond spin chain,
see reference~\cite{effective_xy(b)} and references therein.
As mentioned in section~\ref{sec1},
the magnetic properties of azurite Cu$_3$(CO$_3$)$_2$(OH)$_2$ can be described by a distorted diamond Heisenberg spin chain
with a set of exchange couplings given in equation~(\ref{1.01}) and the gyromagnetic ratio $g=2.06$,
see references~\cite{ah_azurite,effective_xy(a),effective_xy(b)}.
The reduced field $h$ in equation~(\ref{2.01}) or equation~(\ref{2.02}) is related to the physical field ${\cal{H}}$
by $h=g\mu_{\rm{B}}{\cal{H}}$ with $\mu_{\rm{B}}\approx 0.67171$ K/T in the units where $k_{\rm{B}}=1$,
see reference~\cite{effective_xy(a)}.

In what follows we discuss the high-field part of the low-temperature magnetization curve~\cite{kikuchi(a),kikuchi(b)}
which exhibits almost a direct transition
from the plateau at 1/3 of the saturation value to the saturation value
at fields slightly above 30 T and  temperatures about 0.1 K.
This characteristic feature of the magnetization curve may be viewed as a remnant of localized magnons
which dominate the high-field low-temperature properties of the ideal diamond spin chain, see figure~\ref{fig01}.
Herein below we use the approximate analytical description based on equation~(\ref{3.10}).
Another approach to the calculation of magnetization
which is based on variational mean-field-like treatment with the help of Gibbs-Bogolyubov inequality
has been reported recently in reference~\cite{ananikian}.

Bearing in mind the localized-magnon picture emerging for the ideal diamond spin chain,
we may expect for azurite
that the lowest-energy states having different $S^z=N/2,\ldots,N/2-{\cal{N}}$ at a magnetic field around the saturation field,
have almost the same value of energy.
This will obviously produce a very steep part at the low-temperature magnetization around the saturation field.
However, due to a non-ideal geometry, these lowest-energy states are not localized magnons
(yielding a perfect jump in the ground-state magnetization curve),
but almost localized magnons
and their effect on high-field low-temperature thermodynamics can be estimated using equation~(\ref{3.10}).

\begin{figure}[!h]
\begin{center}
\includegraphics[clip=on,width=0.49\textwidth,angle=0]{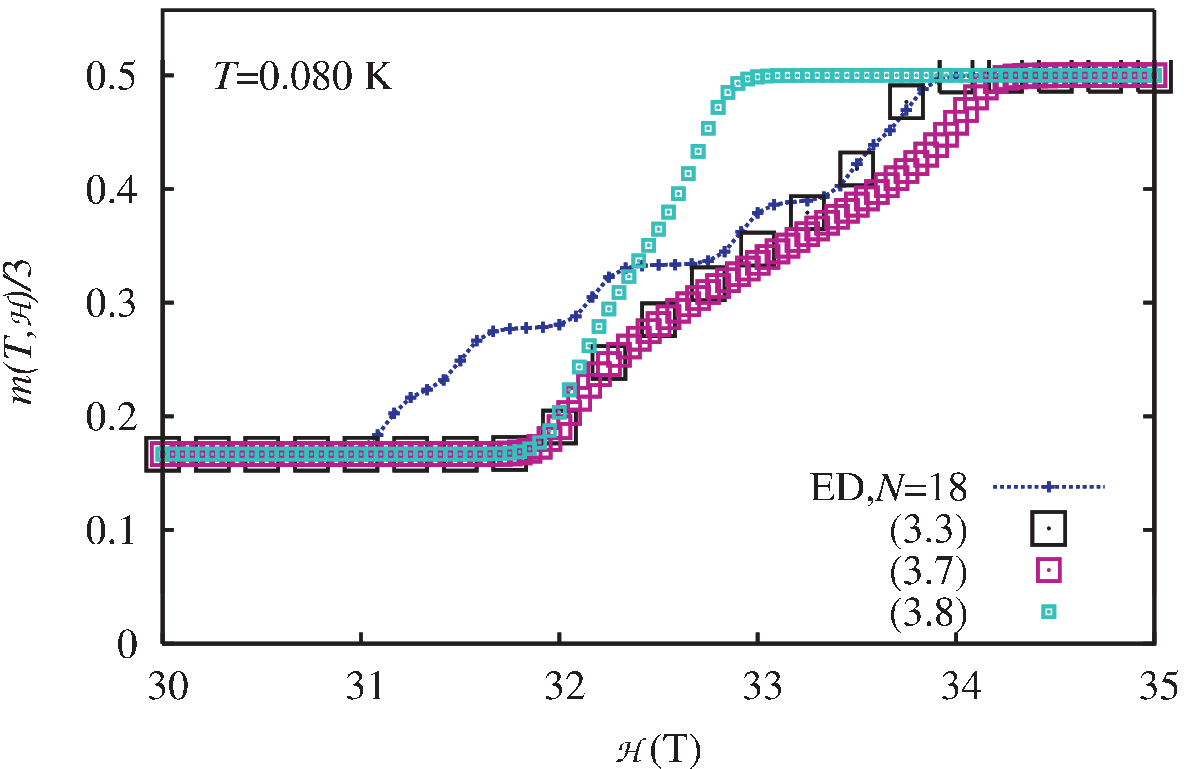}\hfill   
\includegraphics[clip=on,width=0.49\textwidth,angle=0]{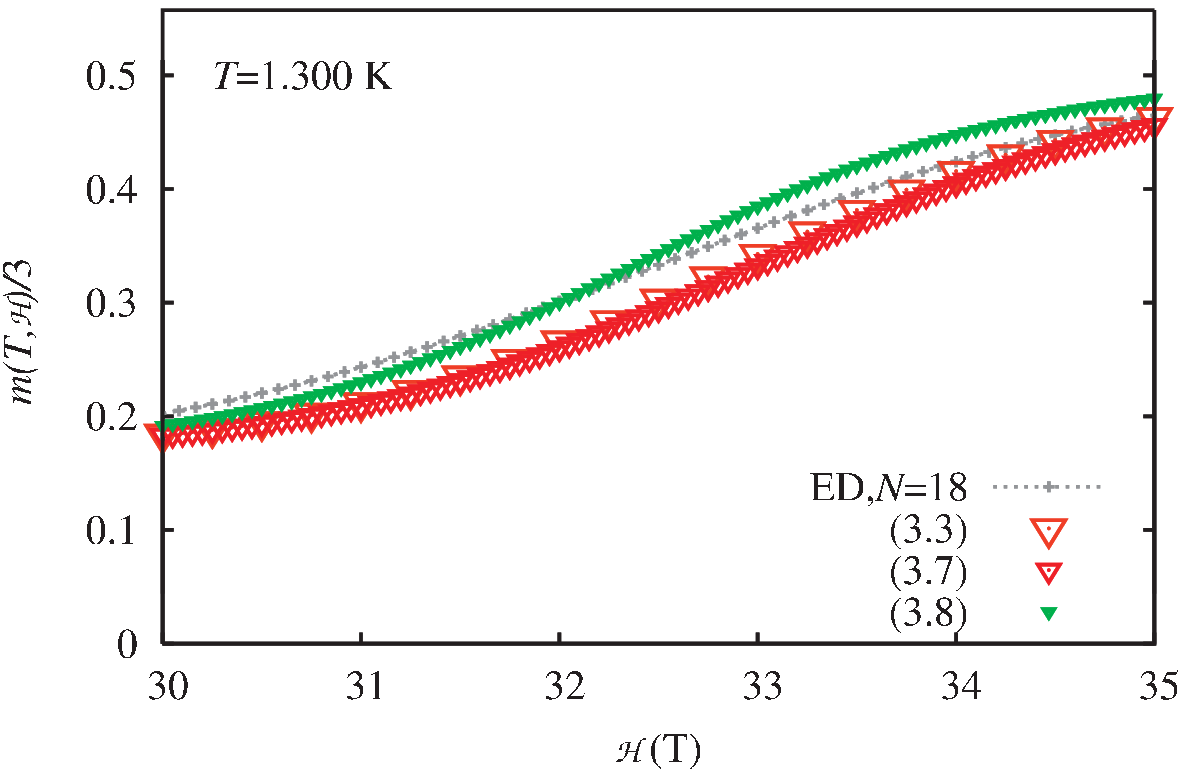}
\caption
{(Color online)
Magnetization curve $m(T,h)/3$ vs ${\cal{H}}$ for the distorted diamond chain
with a set of exchange constants given in equation~(\ref{1.01}) and the value of the gyromagnetic ratio $g=2.06$
at
$T=0.08$ K (left hand panel)
and
$T=1.3$ K (right hand panel).
Exact diagonalization data for $N=18$: lines;
approximate analytical theory which uses equation~(\ref{3.03}): huge empty symbols;
approximate analytical theory which uses equation~(\ref{3.07}): large empty symbols;
approximate analytical theory which uses equation~(\ref{3.08}): small empty symbols.}
\label{fig04}
\end{center}
\end{figure}

Before applying the approach based on equation~(\ref{3.10}) to the model of azurite
we have to make the following remarks.
First of all we note that according to equation~(\ref{1.01}) $(J_1-J_3)/J_2=0.26$,
that is not so small in comparison with the cases reported in figure~\ref{fig03}
(recall we had $\vert J_1-J_3\vert /J_2=0.12,\, 0.10,\, 0.06$ for the upper, middle and lower panels, respectively).
Moreover, now $J_m/J_2=0.14\ne 0$.
Clearly, under these conditions we may question the accuracy of the approach based on equation~(\ref{3.10}).
Anyway,
after inserting the Hamiltonian parameters for azurite
into equation~(\ref{3.15}) with $\varepsilon_1(\kappa)$ following from
equation~(\ref{3.03}) (huge empty symbols),
equation~(\ref{3.07}) (large empty symbols)
or
equation~(\ref{3.08}) (small empty symbols)
we have obtained the results shown in figure~\ref{fig04},
which can be compared with experimental data of reference~\cite{kikuchi(b)},
see figure~2 in this paper.
We also report the corresponding exact diagonalization data for $N=18$.

At the low temperature $T=0.08$ K (left hand panel in figure~\ref{fig04})
the obtained results are in a reasonable agreement with experimental data,
see figure~2 in reference~\cite{kikuchi(b)},
and the exact diagonalization data.
More precisely,
the approximate analytical result which uses $\varepsilon_1(\kappa)$ from equation~(\ref{3.03})
agrees perfectly with the exact diagonalization data above 33.5 T,
though yields the 1/3 plateau already below 32~T, whereas the exact diagonalization prediction is about 31~T. Magnetization  curves with $\varepsilon_1(\kappa)$ from equation~(\ref{3.07}) or equation~(\ref{3.08})
reproduce the exact diagonalization data only qualitatively,
yielding either a larger or a smaller value of the saturation field.
Moreover, slightly above 32~T, approximate analytical results based on equations~(\ref{3.03}), (\ref{3.07}), (\ref{3.08}) coincide.
At the temperature $T=1.3$ K (right hand panel in figure~\ref{fig04})
our calculations show almost no traces of the step-like part nicely seen at $T=0.08$ K (left hand panel in figure~\ref{fig04}).
Furthermore,
at this temperature ($T=1.3$ K) all approximate results are closer to each other and to the exact diagonalization data.

\section{Conclusions}
\label{sec5}
\setcounter{equation}{0}

To summarize,
we have considered the high-field low-temperature properties of a distorted diamond spin chain
using a localized-magnon picture.
The  free energy $f(T,h)$ relevant in this regime is given in equation~(\ref{3.11})
with $\varepsilon_1(\kappa)$ determined from cubic equation (\ref{3.03}),
\begin{eqnarray}
\label{5.01}
&&\varepsilon_i^3(\kappa)+a\varepsilon_i^2(\kappa)+b\varepsilon_i(\kappa)+c=0\,,
\\
&&a=4J+J_2+2J_m\sin^2\frac{\kappa}{2}\,,
\nonumber\\
&&b=-2\left(J^2\cos^2\frac{\kappa}{2}+\delta^2\sin^2\frac{\kappa}{2}\right)
+2(2J+J_2)\left(J+J_m\sin^2\frac{\kappa}{2}\right)+J(J+J_2)\,,
\nonumber\\
&&c=-(2J+J_2)\left(J^2\cos^2\frac{\kappa}{2}+\delta^2\sin^2\frac{\kappa}{2}\right)
-J_2\left(J^2\cos^2\frac{\kappa}{2}-\delta^2\sin^2\frac{\kappa}{2}\right)
\nonumber\\
&&\phantom{c=}+2J(J+J_2)\left(J+J_m\sin^2\frac{\kappa}{2}\right)\nonumber\,,
\end{eqnarray}
i.e.,
\begin{eqnarray}
\label{5.02}
\varepsilon_1(\kappa)=-2\sqrt{-\frac{p}{3}}\cos\frac{\alpha-\pi}{3}-\frac{a}{3}\,,
\;\;\;
p=-\frac{a^2}{3}+b\,,
\;\;\;
q=\frac{2a^3}{27}-\frac{ab}{3}+c\,,
\;\;\;
\cos\alpha=-\frac{q}{2\sqrt{-\frac{p^3}{27}}}\,.
\end{eqnarray}
For small $\delta/J_2$ instead of equation~(\ref{5.02}) we can take $\varepsilon_1(\kappa)$ from  equation~(\ref{3.07}).
Furthermore, assuming that $J/J_2$, $J_m/J_2$ are also small, we arrive at equation~(\ref{3.08}).
Although equation~(\ref{5.02}) for $\varepsilon_1(\kappa)$ provides the best results,
equation~(\ref{3.08}) for $\varepsilon_1(\kappa)$
(valid for $J/J_2\ll1$, $J_m/J_2\ll1$)
permits a very transparent interpretation of the thermodynamics
in terms of the emergent spin-1/2 $XX$ chain in a transverse field (\ref{3.12})
and links our results to the ones obtained earlier within a completely different approach~\cite{effective_xy(a),effective_xy(b)}.

The effective description resembles, to some extent, the spin-1/2 transverse $XX$ chain theory:
Hard-core bosons (spins 1/2) mimic the hard-monomer rule
whereas a nonzero $XX$ exchange coupling is related to a small dispersion of the former flat one-magnon band.
The elaborated description can reproduce the basic features of the high-field magnetization process at low temperatures even quantitatively.
It might be interesting to consider other properties
such as the low-temperature entropy, specific heat or the magnetocaloric effect around the saturation field
within the suggested scheme,
cf. also reference~\cite{effective_xy(b)}.
From reference~\cite{localized_magnons(b)} we know that deviations from ideal geometry
may produce an interesting low-temperature behavior, e.g., of the entropy around the saturation field.

It seems quite evident
that such a description can be applied to other spin systems of the hard-monomer universality class~\cite{localized_magnons(a),localized_magnons(b),localized_magnons(c),localized_magnons(d),fnt},
e.g., the dimer-plaquette chain~\cite{ivanov(a),ivanov(b)} or the two-dimensional square-kagome lattice~\cite{squarekagome(a),squarekagome(b)}.
Another interesting question concerns the applicability of such an approximate approach
to distorted spin systems of other universality classes~\cite{localized_magnons(a),localized_magnons(b),localized_magnons(c),localized_magnons(d),fnt},
in particular, of the hard-dimer universality class.
Moreover,
it might be interesting to use a similar scheme to analyze the high-field low-temperature thermodynamics
of the frustrated triangular spin-tube compound considered in reference~\cite{nedko}
extending the ideal geometry description of references~\cite{maksymenko(a),maksymenko(b)}.
Last but not least,
we may mention a consideration of distorted electron models~\cite{hubbard(a),hubbard(b),hubbard(c),hubbard(d)}.

Finally it should be stressed that in spite the fact that
the suggested approach provides a semiquantitative description
of the high-field low-temperature properties of a distorted diamond spin chain,
it is not clear how it can be systematically improved.
From this point of view, another systematic approach, e.g., a perturbation theory
(though not with respect to small $\vert J_1\vert/J_2$, $\vert J_3\vert/J_2$ but with respect to small $\vert J_1-J_3\vert/J_2$),
is required.
Moreover, to achieve a better agreement with experimental data for azurite,
one apparently has to take the three-dimensional coupling geometry of this compound into account.

\section*{Acknowledgements}

The numerical calculations were performed using J.~Schulenburg's {\it spinpack}~\cite{spinpack}.
The present study was supported by the DFG (project RI615/21-1).
O.~D. acknowledges the kind hospitality of the University of Magdeburg in the spring of 2012.
O.~D. would like to thank the Abdus Salam International Centre for Theoretical Physics (Trieste, Italy)
for partial support of these studies through the Senior Associate award.

\newpage

\ukrainianpart

\title{Напівкількісна теорія низькотемпературних властивостей деформованого ромбічного спінового ланцюжка \\
       у сильних полях}
\author{Олег Держко\refaddr{label1,label2,label3}, Йоганес Ріхтер\refaddr{label3}, Олеся Крупніцька\refaddr{label2}}
\addresses{
\addr{label1} Інститут фізики конденсованих систем НАН України,
              вул. Свєнціцького, 1, 79011 Львів, Україна
\addr{label2} Кафедра теоретичної фізики Львівського національного університету ім. Івана Франка, \\
              вул. Драгоманова, 12,  79005 Львів, Україна
\addr{label3} Інститут теоретичної фізики, Університет Магдебурга,
              D-39016 Магдебург, Німеччина
}
%
%
%

\makeukrtitle

\begin{abstract}
\tolerance=3000%
Ми розглядаємо антиферомагнітну модель Гайзенберга на деформованому ромбічному ланцюжку
і використовуємо картину локалізованих магнонів, пристосовану до деформованої геометрії,
щоб обговорити деякі низькотемпературні властивості моделі у сильних полях.
Конкретніше,
у нашому дослідженні ми вважаємо,
що статистична сума у випадку дещо деформованої геометрії має таку ж форму як і у випадку ідеальної геометрії,
але з трошки дисперсними одномагнонними енергіями.
Ми також обговорюємо застосовність такого опису для азуриту.
\keywords ромбічний спіновий ланцюжок, локалізовані магнони, азурит

\end{abstract}

\end{document}